\documentclass[12pt]{article}
\usepackage{epsf}

\def\hybrid{\topmargin 0pt      \oddsidemargin 0pt
	\headheight 0pt \headsep 0pt
	\textwidth 6.25in       
        \textheight 9.5in       
	\marginparwidth .875in
	\parskip 5pt plus 1pt   \jot = 1.5ex}

\catcode`\@=11
\def\marginnote#1{}

\newcount\hour
\newcount\minute
\newtoks\amorpm
\hour=\time\divide\hour by60
\minute=\time{\multiply\hour by60 \global\advance\minute by-\hour}
\edef\standardtime{{\ifnum\hour<12 \global\amorpm={am}%
	\else\global\amorpm={pm}\advance\hour by-12 \fi
	\ifnum\hour=0 \hour=12 \fi
	\number\hour:\ifnum\minute<10 0\fi\number\minute\the\amorpm}}
\edef\militarytime{\number\hour:\ifnum\minute<10 0\fi\number\minute}

\def\draftlabel#1{{\@bsphack\if@filesw {\let\thepage\relax
   \xdef\@gtempa{\write\@auxout{\string
      \newlabel{#1}{{\@currentlabel}{\thepage}}}}}\@gtempa
   \if@nobreak \ifvmode\nobreak\fi\fi\fi\@esphack}
	\gdef\@eqnlabel{#1}}
\def\@eqnlabel{}
\def\@vacuum{}
\def\draftmarginnote#1{\marginpar{\raggedright\scriptsize\tt#1}}

\def\draft{\oddsidemargin -.5truein
	\def\@oddfoot{\sl preliminary draft \hfil
	\rm\thepage\hfil\sl\today\quad\militarytime}
	\let\@evenfoot\@oddfoot \overfullrule 3pt
	\let\label=\draftlabel
	\let\marginnote=\draftmarginnote
   \def\@eqnnum{(\theequation)\rlap{\kern\marginparsep\tt\@eqnlabel}%
\global\let\@eqnlabel\@vacuum}  }

\def\numberbysection{\@addtoreset{equation}{section}
	\def\theequation{\thesection.\arabic{equation}}}

\catcode`@=12
\relax

\def\beq{\begin{equation}}
\def\eeq{\end{equation}}
\def\bea{\begin{eqnarray}}
\def\eea{\end{eqnarray}}

\relax
\numberbysection
\hybrid

\begin{document}
\begin{titlepage}
\begin{center}
{\large\bf Large-$q$ asymptotics of the random bond Potts model}\\[.3in] 
        {\bf Jesper Lykke Jacobsen%
             \footnote{Present address: LPTMS, b\^{a}t.~100,
             Universit\'e Paris-Sud, F-91405 Orsay, France.} 
             (1) and Marco Picco (2)} \\ 
        {\bf (1)} {\it Laboratoire de Physique Statistique%
             \footnote{Laboratoire associ{\'e} aux universit{\'e}s
                       Paris 6, Paris 7 et au CNRS.},\\
             Ecole Normale Sup{\'e}rieure,\\
             24 rue Lhomond,
             F-75231 Paris CEDEX 05, FRANCE \\}
        {\bf (2)} {\it LPTHE\/}\footnote{Unit{\'e} Mixte de Recherche CNRS
                   UMR 7589.},\\
        {\it  Universit{\'e} Pierre et Marie Curie, PARIS VI\\
              Universit{\'e} Denis Diderot, PARIS VII\\
              Bo\^{\i}te 126, Tour 16, 1$^{\it er}$ {\'e}tage \\
              4 place Jussieu,
              F-75252 Paris CEDEX 05, FRANCE}\\

\end{center}
\vskip .15in
\centerline{\bf ABSTRACT}
\begin{quotation}

{\small We numerically examine the large-$q$ asymptotics of the
$q$-state random bond Potts model. Special attention is paid to the
parametrisation of the critical line, which is determined by combining
the loop representation of the transfer matrix with Zamolodchikov's
$c$-theorem. Asymptotically the central charge seems to behave like $c(q) =
\frac12 \log_2(q) + {\cal O}(1)$. Very 
accurate values of the bulk magnetic exponent $x_1$ are then extracted by
performing Monte Carlo simulations directly at the critical point. 
As $q\to\infty$, these seem to tend to a non-trivial limit,
$x_1 \to 0.192 \pm 0.002$.}



\end{quotation}
\end{titlepage}

\section{Introduction}

Recently the two-dimensional $q$-state random bond Potts model
with $q>4$ has attracted considerable interest, because it serves as a
paradigm for examining the effect of quenched randomness
\cite{Cardy-conf} on a first-order phase transition \cite{Baxter73}.
Since in this case the randomness couples to
the local energy density, a theorem by Aizenman and Wehr \cite{Aizenman-Wehr},
along with related analytical work \cite{Imry-Wortis,Hui-Berker},
suggests that the transition should become continuous, as has indeed
been verified by subsequent numerical studies
\cite{Chen95,Picco97,Cardy97,Cardy98,Picco98,Berche,Olson-Young,Palagyi}.
Unfortunately, analytical results have been scarce, except in the limit
$q\to\infty$ where properties of a particular tricritical point were
related to those of the zero-temperature fixed point of the random
{\em field} Ising model in $d=2+\varepsilon$ dimensions \cite{Cardy97}.
{}From the conjectured phase
diagram \cite{Cardy97} it is however known that this fixed point is
not the analytical continuation of the line of random fixed points
found for finite $q>2$ \cite{Ludwig-Cardy,Ludwig}. Namely the latter
(henceforth referred to as the $q\to\infty$ limit of the model) is
rather believed to be associated with a subtle percolation-like limit
\cite{Cardy97}, the exact properties of which have not yet been fully
elucidated.

In the present publication we seek to gain further knowledge of this
$q\to\infty$ limit by producing numerical results along the
afore-mentioned line of critical fixed points for very large values of
$q$. Since cross-over effects to the pure and percolative limits of
the model have been shown to be important \cite{Cardy98,Picco98},
special attention must be paid to the parametrisation of the critical
line. Generalising 
a recently developed transfer matrix technique \cite{Dotsenko99}, in which
the Potts model is treated through its loop representation
\cite{Baxter82}, we were able to explicitly trace out this line, and as a
by-product obtain very precise values of the central charge. Based on
our numerical results for the $q=8^k$ state model with
$k=1,2,\ldots,6$ we find compelling evidence that
\beq
  c(q) = \frac12 \log_2(q) + {\cal O}(1).
  \label{cIsing}
\eeq
Although this behaviour of the central charge is reminiscent of the
Ising-like features of the tricritical fixed point discussed above, we
shall soon see that from the point of view of the magnetic exponent the
$q\to\infty$ limit is most definitely not in the Ising universality class.
Note also that our precision allows us, for the first time, to
convincingly distinguish the numerically computed central charge from
its analytically known value in the percolation limit \cite{Cardy98}.

With the numerically obtained parametrisation of the critical disorder
strength at hand we then proceed to measure the corresponding magnetic bulk
scaling dimension $x_1$ as a function of $q$. The most suitable
technique is here that of conventional Monte Carlo simulations. Our
results lend credibility to the belief \cite{Olson-Young} that
$x_1(q)$ saturates as $q\to\infty$. Based on results for the $q=8^k$
state model with $k=1,2,3$ we propose the limiting value
\beq
  x_1(q) \to 0.192 \pm 0.002 \ \ \mbox{\rm for} \ \ q\to\infty,
  \label{x1infty}
\eeq
in agreement with the one reported in Ref.~\cite{Olson-Young}.
The fact that Eq.~(\ref{x1infty}) does not coincide with any known scaling
dimension of standard percolation is remarkable, and calls for further
analytical investigations of the $q\to\infty$ limit.

After explaining the loop model transfer matrices in
Section~\ref{sec:loop}, we state our results for the critical line and
the central charge in Section~\ref{sec:cc}. The Monte Carlo method and
the resulting values of the magnetic scaling dimension are presented in
Section~\ref{sec:mc}, and we conclude with a discussion.

\section{Loop model transfer matrices}
\label{sec:loop}

The partition function of the random bond Potts model can be written
as 
\beq
 Z = \sum_{\{\sigma\}} \prod_{\langle ij \rangle}
     {\rm e}^{K_{ij} \delta_{\sigma_i,\sigma_j}},
 \label{Z1}
\eeq
where the summation is over the $q$ discrete values of each spin and
the product runs over all nearest-neighbour bonds on the square lattice.
The $K_{ij}$ are the reduced coupling constants, which for the moment
may be drawn from an arbitrary distribution.
By the standard Kasteleyn-Fortuin transformation \cite{Kasteleyn},
Eq.~(\ref{Z1}) can be recast as a random cluster model
\beq
 Z = \sum_{\{{\cal G}\}} q^{C({\cal G})}
     \prod_{\langle ij \rangle \in {\cal G}}
     \left({\rm e}^{K_{ij}}-1 \right),
 \label{Z2}
\eeq
where ${\cal G}$ is a bond percolation graph with $C({\cal G})$
independent clusters. Note that $q$ now enters only as a (continuous)
parameter, and since the non-locality of the clusters does not
obstruct the construction of a transfer matrix \cite{Blote82} the
interesting regime of $q \gg 4$ becomes readily accessible, provided
that one can take into account the randomness in the couplings
\cite{Cardy98}.

In an analogous fashion we can adapt the even more efficient loop
model representation \cite{Dotsenko99} to the random case. Indeed,
trading the clusters for their surrounding loops on the medial lattice
\cite{Baxter82}, Eq.~(\ref{Z2}) is turned into
\beq
  Z = q^{N/2} \sum_{\{{\cal G}\}} q^{L({\cal G})/2}
     \prod_{\langle ij \rangle \in {\cal G}}
     \left( \frac{{\rm e}^{K_{ij}}-1}{\sqrt{q}} \right),
\eeq
where $N$ is the total number of spins, and configuration ${\cal G}$
encompasses $L({\cal G})$ loops. The strip width $L$ is measured in
terms of the number of `dangling' loop segments, and must be even by
definition of the medial lattice \cite{Dotsenko99}.

A pleasant feature of the random bond Potts model is that the
critical temperature is known exactly by self-duality \cite{Kinzel}.
Employing for simplicity the bimodal distribution
\beq
 P(K_{ij}) = \frac12 \left[ \delta(K_{ij}-K_1) + \delta(K_{ij}-K_2) \right],
 \label{dist}
\eeq
and choosing the parametrisation
$s_{ij} \equiv ({\rm e}^{K_{ij}}-1)/\sqrt{q}$,
the self-duality criterion takes the simple form
\beq
 s_1 s_2 = 1.
\eeq
To fully identify the critical point the only free parameter is then
the strength of the disorder, which can be measured in terms of
$R \equiv K_1/K_2 > 1$ or $s \equiv s_1 > 1$.

\section{Central charge}
\label{sec:cc}

In Ref.~\cite{Dotsenko99} we showed that Zamolodchikov's $c$-theorem
\cite{Zamolodchikov} is a powerful tool for numerically identifying
the fixed points of a {\em pure} system. The idea is simple: {}From the
leading eigenvalue of the transfer matrix, specific free energies
$f_0(L)$ can be computed as a function of the strip width $L$. Effective
central charges $c(L)$ are then obtained by fitting data for two
consecutive strip widths according to \cite{Cardy86}
\beq
  f_0(L) = f_0(\infty) - \frac{\pi c}{6 L^2} + \cdots.
  \label{f0}
\eeq
By tuning the free parameter $s$ of the system, local extrema
$c(L,s_*(L))$ are sought for, and finally the fixed point is
identified by extrapolation: $s_* = s_*(L\to\infty)$.

In principle this strategy can also be employed for a {\em disordered}
system, provided that error bars are carefully kept under control. 
Now $f_0(L)$ is related to the largest Lyapunov exponent of a product
of $M\to\infty$ random transfer matrices \cite{Furstenberg,Cardy97},
and its statistical error vanishes as $M^{-1/2}$ by the central limit
theorem. Thus, for large enough $M$ any desired precision on $f_0(L)$
can be achieved.

An important observation is that for larger and larger $L$, the $c(L)$
found from Eq.~(\ref{f0}) become increasingly sensitive to errors in
$f_0(L)$. Therefore $M$ must be chosen in accordance with the largest
strip width $L_{\rm max}$ used in the simulations. For the system at
hand we found that 4 significant digits in $c(L)$ were needed for a
reasonable precise identification of $s_*(L)$, and with
$L_{\rm max}=12$ this in turn implies that the $f_0(L)$ must be
determined with 6 significant digits. We were thus led to choose
$M=10^8$ for $q=8$, and $M=10^9$ for larger values of $q$.%
\footnote{Incidentally, improving our results to $L_{\rm max}=14$
would require augmenting $M$ by at least a factor of 100 (apart from
the increased size of the transfer matrices), and since several months
of computations were spent on the present project this hardly seems
possible in a foreseeable future.}

Data collection was done by dividing the strip into $M/l$ patches of
length $l=10^5$ lattice spacings, and for each patch the couplings were
randomly generated from a {\em canonical} ensemble, i.e., the
distribution (\ref{dist}) was restricted to produce an equal number of
strong and weak bonds.

\begin{table}
 \begin{center}
 \begin{tabular}{r|ccc|cccc}
 $s$ & $c(4,8)$ & $c(6,10)$ & $c(8,12)$ &
       $c(4,6)$ & $c(6,8)$  & $c(8,10)$ & $c(10,12)$ \\ \hline
   3 & 1.495 & 1.500 & 1.500 & 1.4101 & 1.4544 & 1.4731 & 1.4821 \\
   4 & 1.512 & 1.517 & 1.516 & 1.4157 & 1.4657 & 1.4868 & 1.4967 \\
   5 & 1.519 & 1.525 & 1.523 & 1.4152 & 1.4690 & 1.4918 & 1.5025 \\
   6 & 1.521 & 1.528 & 1.527 & 1.4116 & 1.4683 & 1.4927 & 1.5044 \\
   7 & 1.520 & 1.529 & 1.529 & 1.4067 & 1.4656 & 1.4915 & 1.5041 \\
   8 & 1.518 & 1.528 & 1.529 & 1.4013 & 1.4619 & 1.4890 & 1.5026 \\
   9 & 1.509 & 1.530 & 1.528 & 1.3972 & 1.4552 & 1.4860 & 1.5004 \\
  10 & 1.511 & 1.525 & 1.527 & 1.3908 & 1.4534 & 1.4826 & 1.4977 \\
  11 & 1.501 & 1.526 & 1.526 & 1.3873 & 1.4465 & 1.4791 & 1.4949 \\
  12 & 1.504 & 1.519 & 1.524 & 1.3816 & 1.4451 & 1.4756 & 1.4919 \\
 \end{tabular}
 \end{center}
 \protect\caption[2]{\label{tab:cc}Effective central charge of the
 $q=8$ state model, as a function of disorder strength $s$.
 Two- and three-point fits to Eq.~(\ref{f0}) are
 labelled as $c(L,L+2)$ and $c(L,L+4)$ respectively.}
\end{table}

In the right part of Table~\ref{tab:cc} we show the resulting
two-point fits (\ref{f0}) in the $q=8$ state model, as a function of
$s$. The left part of the table provides analogous three-point fits,
obtained by including a non-universal $1/L^4$ correction in
Eq.~(\ref{f0}). In all cases the error bars are believed to affect
only the last digit reported. The two-point fits give clear evidence
of a maximum in the central charge,
and we estimate its location as $s_* = 6.5 \pm 2.0$.
The corresponding central charge is estimated from the three-point
fits, as these are known to converge faster in the $L\to\infty$ limit
\cite{Cardy98}, and we arrive at $c=1.530 \pm 0.001$.
To appreciate the precision of this result, we mention that the
numerical values of $c(q=8)$ first reported were
$1.50 \pm 0.05$ \cite{Picco97} and $1.517 \pm 0.025$ \cite{Cardy97}.

Table~\ref{tab:sc} summarises our results for other values of $q$.
Two remarkable features are apparent. First, $s_* \propto q^w$ is well
fitted by a power law with $w=0.31 \pm 0.02$. This gives valuable
information on how the $q\to\infty$ limit of the model is approached,
and implies that the ratio of the coupling constants
$R \equiv K_1/K_2 = \log(1+s \sqrt{q}) / \log(1+\sqrt{q}/s)$ is a
non-monotonic function of $q$ that tends to the {\em finite} limiting value
$\frac{1+2w}{1-2w} = 4.3 \pm 0.6$ as $q\to\infty$.
We shall discuss this finding further in Section~\ref{sec:disc}.
Second, the central charge seems to fulfil the relation (\ref{cIsing})
as stated in the Introduction.

\begin{table}
 \begin{center}
 \begin{tabular}{r|ccc}
  $q$   & $s_*$     & $c$        & $c/\log_2(q)$ \\ \hline
      8 & 6.5 (20)  & 1.530 (1)  & 0.5100 (3)    \\
     64 & 15.5 (20) & 3.050 (3)  & 0.5083 (5)    \\
    512 & 32 (2)    & 4.545 (10) & 0.5050 (11)   \\
   4096 & 65 (8)    & 6.038 (24) & 0.5032 (20)   \\
  32768 & 135 (20)  & 7.54 (3)   & 0.5027 (20)   \\
 262144 & 250 (50)  & 9.04 (3)   & 0.502 (2)     \\
 \end{tabular}
 \end{center}
 \protect\caption[2]{\label{tab:sc}Critical disorder strength $s$ and
 central charge $c$, as functions of $q$.}
\end{table}

\section{Magnetic scaling dimension}
\label{sec:mc}

In this section we explain the Monte Carlo method used for obtaining
values of the magnetic scaling exponent. Simulations were performed on
square lattices of size $L\times L$ with periodic boundary conditions,
with $L$ ranging from 4 to $L_{\rm max}=128$ for $q=8,64$ and
$L_{\rm max}=64$ for $q=512$.

We employed the Wolff cluster algorithm \cite{wolff}. The first part
of the simulations was to determine the autocorrelation times $\tau$,
which were found to increase with the lattice size and also with
$q$. For the largest simulated lattices, we determined $\tau$ as follows:
$88 \pm 4$ cluster updates for $q=8$ and $L=128$, $3000 \pm 215$ for
$q=64$ and $L=128$, and $31000 \pm 3000$ for $q=512$ and $L=64$.
This rapid increase of $\tau$ with $q$ explains why we simulate
only up to $L=64$ for the largest $q$. 

Next, we measure the magnetisation, defined for each disorder sample
$x$ by 
\beq
m_x = {{q \langle \rho \rangle -1}\over{q-1}},
\eeq
where $\rho=\max(N_1,N_2,\ldots,N_q)/L^2$ and $N_\sigma$ is the number of
Potts spins taking the value $\sigma$. Here $\langle \ldots \rangle$
denotes the thermal average. Then the magnetisation $m(L)$ is obtained by
averaging over $10^5$ disorder configurations for $q=8$, and
$10^4$ configurations for $q=64$ and $512$. For each disorder
sample, $100 \times \tau$ updates were dedicated to the thermalisation,
and a further $100 \times \tau$ to the magnetisation measurement.
Error bars were computed from the disorder fluctuations (it can easily
be checked \cite{Picco98} that the contribution from {\em thermal}
fluctuations is negligible), and the strength of the disorder was
chosen as indicated in Table \ref{tab:sc}.

{}From a fit to $m(L) \simeq L^{-x_1}$, we obtain
for the magnetic scaling dimension:
\beq
x_1 = \left \lbrace \begin{array}{lll}
 0.1535(10) & \mbox{for} & q=8 \\
 0.172(2)   & \mbox{for} & q=64 \\
 0.180(3)   & \mbox{for} & q=512 \\
 \end{array} \right.
\eeq

We see that the magnetic exponent seems to saturate as we increase
$q$. In view of the result (\ref{cIsing}) for the central charge we
expect the asymptotic behaviour should involve $\log(q)$ rather than
$q$ itself, and indeed the data are well fitted by
\beq
 x_1(q) = a + b / \log(q)
 \label{form}
\eeq
with $a=0.192(2)$ and $b=-0.080(4)$. Thus, based on the form
(\ref{form}) we are led to propose the limiting value (\ref{x1infty})
of $x_1$ given in the Introduction.

\section{Discussion}
\label{sec:disc}

It is useful to juxtapose our findings on the large-$q$ behaviour of the
critical line with the phase diagram proposed in Ref.~\cite{Cardy97}. In
that work the disorder strength was parametrised through $s=q^w$ with
$w>0$, and the limit $w\to\infty$ was identified with classical
percolation on top of the strong bonds. Actually it is easily seen
from Eq.~(\ref{Z2}) that directly at $q=\infty$ this percolation
scenario holds true whenever $w > 1/2$, and assuming that the line of
critical fixed points is described by a monotonic function $w_*(q)$ it
can thus be confined to the region $w \le 1/2$.
With this slight reinterpretation, Ref.~\cite{Cardy97} argues that at
$q=\infty$ the critical point is located in the limit $w \to 1/2$.
Indeed, since for $q=\infty$ any initial $w \ll 1/2$ will be
driven to larger values due to mapping to the random field Ising
model, this is nothing but the usual assumption of ``no intervening
fixed points''.

However, this seems at odds with the results of Table~\ref{tab:sc},
where we found that for $q \gg 4$ the critical line, when measured in
terms of $w$, saturates at $w=0.31 \pm 0.02$. Unless our numerical
method is flawed by some gross systematic error, it is thus a priori
difficult to see how this can be reconciled with the above result of
$w_*(q=\infty) = 1/2$. A possible explanation is that 
the limits $q\to\infty$ and $w\to 1/2$ are highly
non-commuting. This is witnessed by the jump in the central charge,
which in the percolation limit ($w=\infty$ and $q<\infty$) reads
\cite{Cardy97}
\beq
 c_{\rm perc} = \frac{5 \sqrt{3}}{4 \pi} \ln(q) \simeq 0.47769 \log_2(q),
\eeq
to be contrasted with our numerical result (\ref{cIsing}).

\noindent{\large\bf Acknowledgments}

We are grateful to J.~Cardy for some very helpful comments.

\end{document}